\documentclass[twocolumn,journal]{IEEEtran}

\usepackage{amsfonts, amssymb, amsmath, cite, enumerate, amsthm, bm, caption, subfig, psfrag, cases, mathtools}
\usepackage{algorithm}
\usepackage{booktabs}
\usepackage{algorithmic}
\usepackage{ifthen}
\usepackage[usenames,dvipsnames]{color}

\ifCLASSINFOpdf
   \usepackage[pdftex]{graphicx}
   \DeclareGraphicsExtensions{.pdf,.jpeg,.png}
\else
   \usepackage[dvips]{graphicx}
   \DeclareGraphicsExtensions{.eps}
\fi

\DeclareMathOperator{\sign}{sgn}

\newcommand{\ip}[2]{\left\langle#1,#2\right\rangle}
\newcommand{\norm}[1]{\left\lVert#1\right\rVert}
\renewcommand{\(}{\left(}
\renewcommand{\)}{\right)}
\renewcommand{\[}{\left[}
\renewcommand{\]}{\right]}

\newboolean{draft}
\newcommand{\isdraft}[2]{\ifthenelse{\boolean{draft}}{#1}{#2}}

\isdraft{\usepackage{setspace}}{}                              
\isdraft{\usepackage[footnotesize]{caption}}{}                 
\isdraft{\usepackage{paralist}}{}                              

\def \C {\mathbb{C}}

\def \R {\mathbb{R}}

\def \< {\langle}
\def \> {\rangle}

\def \ve {\bm{e}}
\def \vf {\bm{f}}

\def \vh {\bm{h}}
\def \vj {\bm{j}}

\def \vn {\bm{n}}

\def \vx {\bm{x}}

\def \vy {\bm{y}}
\def \vz {\bm{z}}
\def \vX {\bm{X}}
\def \vY {\bm{Y}}
\def \vA {\bm{A}}
\def \vB {\bm{B}}

\def \vF {\bm{F}}

\def \vN {\bm{N}}

\def \vW {\bm{W}}

\def \vI {\bm{I}}
\def \vV {\bm{V}}
\def \vU {\bm{U}}

\def \vZ {\bm{Z}}

\def \vphi {\bm{\phi}}

\def \vSigma {\bm{\Sigma}}

\def \vLambda {\bm{\Lambda}}

\def \vPhi    {\bm{\Phi}}

\theoremstyle{plain}
\newtheorem{theorem}{Theorem}

\newtheorem{corollary}{Corollary}

\newtheorem{lemma}{Lemma}

\theoremstyle{remark}
\newtheorem{remark}{Remark}

\begin{document}
\title{Spectrally Sparse Signal Recovery via Hankel Matrix Completion with Prior Information}

\author{Xu~Zhang, Yulong~Liu and Wei~Cui
	\thanks{X.~Zhang and W.~Cui are with the School of Information and Electronics, Beijing Institute of Technology, Beijing 100081, China (e-mail: connorzx@bit.edu.cn; cuiwei@bit.edu.cn).}
	\thanks{Y.~Liu is with the School of Physics,
		Beijing Institute of Technology, Beijing 100081, China (e-mail: yulongliu@bit.edu.cn).}
}%

%



\maketitle

\pagestyle{empty}  
\thispagestyle{empty} 

\begin{abstract} 
This paper studies the problem of reconstructing spectrally sparse signals from a small random subset of time domain samples via low-rank Hankel matrix completion with the aid of prior information. By leveraging the low-rank structure of spectrally sparse signals in the lifting domain and the similarity between the signals and their prior information, we propose a convex method to recover the undersampled spectrally sparse signals. The proposed approach integrates the inner product of the desired signal and its prior information in the lift domain into vanilla Hankel matrix completion, which maximizes the correlation between the signals and their prior information.
Theoretical analysis indicates that when the prior information is reliable, the proposed method has a better performance than vanilla Hankel matrix completion, which reduces the number of measurements by a logarithmic factor. We also develop an ADMM algorithm to solve the corresponding optimization problem. Numerical results are provided to verify the performance of proposed method and corresponding algorithm.

\end{abstract}
%

\section{Introduction}

Spectrally sparse signal recovery refers to recovering a spectrally sparse signal from a small number of time domain samples, which is fundamental in various applications, such as medical imaging\cite{lustig2007sparse}, radar imaging\cite{potter2010sparsity}, analog-to-digital conversion\cite{tropp2009beyond} and channel estimation\cite{paredes2007ultra}. 
Let $\vx=[x_0,\ldots,x_{n-1}]^T \in \C^n$ denote the one-dimensional spectrally sparse signal to be estimated.  Each entry of the desired signal $\vx$ is a weighted superposition of $r$ complex sinusoids
$$
x_k=\sum_{l=1}^{r}w_l e^{i2\pi k f_l}, 
$$
where $k=0,\ldots,n-1$, $\{f_1,\ldots,f_r\}$ and $\{w_1,\ldots,w_r\}$ denote the normalized frequencies and amplitudes for the $r$ sinusoids, respectively, and $f_l \in [0,1)$ for $l=1,\ldots,r$.  

In many practical applications, we only have access to a small subset of signal samples. For example, in the field of computed tomography (CT),  only part of the desired signals can be observed to protect the patients from high-dose radiation\cite{brenner2007computed}; in wideband signal sampling, it's challenging to build analog-to-digital converter according to Shannon sampling theorem, and hence only a small number of samples of the wideband signals can be acquired for reconstruction\cite{tropp2009beyond}.
Therefore, we have to figure out a way to recover the original  signal $\vx$ from its random undersampled observations
$$\mathcal{P}_\Omega(\vx)=\sum_{k \in \Omega} x_k \ve_k,$$
where $\Omega \in \{0,\ldots,n-1\}$ denote the index set of the entries we observe, $\ve_k$ denotes the $k$-th canonical basis of $\R^n$, and $\mathcal{P}_\Omega(\cdot)$ denotes the projection operator on the sampling index set $\Omega$, i.e., $\mathcal{P}_\Omega(\vz)=\sum_{k \in \Omega} \ip{\vz}{\ve_k} \ve_k$ for $\vz \in \C^n$.

In order to reconstruct $\vx$, structured low-rank completion methods have been proposed by using the low-rank Hankel structure of the spectrally sparse signals in the lifting domain
\begin{equation} \label{P: low-rank matrix completion}
\begin{aligned}
&\min \limits_{\vz} ~~ \text{Rank}(\mathcal{H}(\vz)) \\
&~\text{s.t.}~~~  \mathcal{P}_\Omega(\vz)= \mathcal{P}_\Omega({\vx}),
\end{aligned}	
\end{equation}
where $\text{Rank}(\cdot)$ returns the rank of matrix, and $\mathcal{H}: \C^n \to \C^{n_1 \times n_2}$ is a linear lifting operator to generate the Hankel low-rank structure. In particular, for a vector $\vx=[x_0, x_1,\ldots, x_{n-1}]^T \in \C^n$, the Hankel matrix $\mathcal{H}(\vx)$ is defined as
$$\mathcal{H}(\vx) \triangleq \left[ \begin{array}{cccc}
{x_0} & {x_1} & {\ldots} & {x_{d-1}}  \\ 
{x_1} & {x_2} & {\ldots} & {x_d}  \\
{\vdots} & {\vdots} & {\ddots} & {\vdots}  \\
{x_{n-d}} & {x_{n-d+1}} & {\ldots} & {x_{n-1}}  \\		
\end{array} \right], $$
where $d$ denotes the matrix pencil parameter, $n_1=n-d+1$ and $n_2=d$. 


By using Vandermonde decomposition, the Hankel matrix $\mathcal{H}(\vx)$ can be decomposed as
$$
   \mathcal{H}(\vx)=\sum_{l=1}^{r} w_l \vy_l \vz_l^H,
$$ 
where $\vy_l=[1,e^{i2\pi f_l},\ldots,e^{i2\pi (n_1-1) f_l}]^T$ and $\vz_l=[1,e^{i2\pi f_l},\ldots,e^{i2\pi (n_2-1) f_l}]^T$, $l=1,\ldots,r$. When the frequencies are all distinct and $r \ll \min\{n_1,n_2\}$, $\mathcal{H}(\vx)$ is a low-rank matrix with $\text{Rank}(\mathcal{H}(\vx))\le r$. 


Since Eq. \eqref{P: low-rank matrix completion} is a non-convex problem and solving it is  NP-hard,   an alternative approach based on convex relaxation is proposed to complete the low rank matrix, that is, Hankel matrix completion program 
\cite{chen2014robust,ye2017compressive}
\begin{equation} \label{P: nuclear norm minimization}
\begin{aligned} 
&\min \limits_{\vz} ~~ \norm{\mathcal{H}(\vz)}_* \\\
&~\text{s.t.}~~~  \mathcal{P}_\Omega(\vz)= \mathcal{P}_\Omega(\vx),
\end{aligned} 	
\end{equation}
where $\norm{\cdot}_*$ denotes the nuclear norm. Theoretical analysis was given to show that $O(r \log^4n)$ samples are enough to recover the desired signal with high probability \cite{chen2014robust}.

Apart from the sparsity constraint in the spectral domain, a reference signal $\vphi$ that is similar to the original signal $\vx$ sometimes is available to us. There are two main sources of this kind of prior information. The first source comes from natural (non-self-constructed) signals. In high resolution MRI \cite{haldar2008anatomically,peng2011reference,wu2011prior}, adjacent slices show good similarity with each other; in multiple-contrast MRI  \cite{bilgic2011multi,qu2014magnetic,huang2014fast}, different contrasts in the same scan are similar in structure; in dynamic CT \cite{selim2016low}, the scans for the same slice in different time have similar characteristics. The second source comes from self-constructed signals. One way is to use classical method to construct a similar signal. For example, filtered backprojection reconstruction algorithm from the dynamic scans was used to construct the prior information in dynamic CT \cite{chen2008prior}; smooth method was used to generate prior information in sparse-view CT \cite{selim2018image}; the standard spectrum of dot object was used as the prior information in radar imaging \cite{li2019spectrum}. The other way is to use machine learning to generate a similar signal. In \cite{vella2019single}, the authors generated the reference image by using a CNN network; similarly, other algorithms from deep learning can be used to create reference signals, see e.g. in \cite{dong2015image,kim2016accurate}.

In this paper, we propose a convex approach to integrate  prior information into the reconstruction of spectrally sparse signals by maximizing the correlation between signal $\vz$ and prior information $\vphi$ in the lifting domain
\begin{equation}\label{P: nuclear norm minimization with MC}
\begin{aligned} 
&\min \limits_{\vz} ~~ \norm{\mathcal{H}(\vz)}_*-2 \lambda \, \text{Re}\(\ip{{\mathcal{G}(\vphi)}}{\mathcal{H}(\vz)}\) \\
&~\text{s.t.}~~~  \mathcal{P}_\Omega(\vz)= \mathcal{P}_\Omega(\vx),
\end{aligned}
\end{equation}
where $\lambda>0$ is a tradeoff parameter, $\mathcal{G}(\cdot)=\mathcal{\mathcal{F}(\mathcal{H}(\cdot))}$ is a composition operator, $\ip{\vX}{\vY}=(\mbox{vec}(\vY))^H \mbox{vec}(\vX)$ is the inner product  and $\text{Re}(\cdot)$ returns the real part of a complex number. Here, $\mathcal{F}:\C^{n_1 \times n_2} \to \C^{n_1 \times n_2}$ is a suitable operator to be designed in the sequel.
Theoretical guarantees are provided to show that our method has better performance than vanilla Hankel matrix completion when the prior information is reliable. In addition, we propose an Alternating Direction Method of Multipliers (ADMM)-based optimization algorithm for efficient reconstruction of the desired signals.


\subsection{Related Literature}

Recovery of spectrally sparse signals  has attracted great attentions in the past years. Conventional compressed sensing \cite{candes2006robust,donoho2006compressed} was used to estimate the spectrally sparse signals when the frequencies are located on a grid.  In many practical applications, however, the frequencies lie off the grid, leading to the mismatch for conventional compressed sensing.

To recover the signals with off-the-grid frequencies, two kinds of methods are proposed: atomic norm minimization and low-rank structured matrix completion. By promoting the sparsity in a continuous frequency domain, atomic norm minimization \cite{tang2013compressed,tang2014near}  demonstrated that $r\log(r)\log(n)$ random samples are sufficient to recover the desired signals exactly with high probability when the frequencies are well separated. Due to the fact that the sparsity in frequency domain leads to the low-rankness in the lifting time domain, low rank structured matrix completion \cite{chen2014robust,ye2017compressive} was proposed to promote the low-rank structure in the lifting time domain. Their results  showed that $O(r\log^4(n))$ random samples are enough to correctly estimate the original signals with high probability when some incoherence conditions are satisfied.

Besides the sparse prior knowledge, other kinds of prior information are used to further improve the recovery performance. By using the similarity between original signal and reference signal, an adaptive weighted compressed sensing approach was considered in \cite{weizman2016reference}, which presented a better performance than conventional approach. Assuming that some frequency intervals or likelihood of each frequency of the desired signal is known a priori, a weighted atomic norm method was studied in \cite{mishra2015spectral,li2019compressive}, which outperforms standard atomic norm approach. 

While the above work considered spectrally sparse signal recovery with prior information based on conventional compressed sensing or atomic norm minimization, little work incorporates the prior information into low-rank structured matrix completion.

Recently, we proposed a novel method to recover structured signals by using the prior information via maximizing correlation \cite{zhang2018recovery,Zhang2020Matrix}. By introducing a negative inner product of the prior information and the desired signal into the objective function, theoretical guarantees and numerical results illustrated that the matrix completion approach proposed in \cite{Zhang2020Matrix}  outperforms standard matrix completion procedure in \cite{fazel2002matrix,recht2010guaranteed,candes2009exact,gross2011recovering} when the prior information is reliable.

Inspired by \cite{Zhang2020Matrix}, this paper leverages the transform low-rank information in the lifting domain to recover the undersampled spectrally sparse signals with the help of the prior information. 
Different from \cite{Zhang2020Matrix}, this paper studies the low-rank property in the lifting domain while the previous approach studies the low-rank property in original domain, leading to the change of the desired matrix from random matrix to Hankel random matrix. Accordingly, the sampling operator changes from sampling random entries to sampling random  skew-diagonal. Therefore, different theoretical guarantees should be given to analyze the proposed approach. The analysis also should be extended from real number domain to complex number domain since the spectrally sparse signals are complex.

\subsection{Paper Organization}
The structures of this paper are arranged as follows.  Preliminaries are provided in Section \ref{sec: Preliminaries}. Performance guarantees are given in Section \ref{sec: Performance guarantees}. An extension to multi-dimensional models is provided in Section \ref{sec: multi-dimensional models}. The ADMM optimization algorithm is presented in Section \ref{sec: Optimization Algorithm}.  Simulations are included in Section \ref{sec: Simulation}, and the conclusion is drawn in Section \ref{sec: Conclusion}.

\section{Preliminaries} \label{sec: Preliminaries}

In this section, we introduce some important notation and definitions, which will be used in the sequel. 

 Let $\{\vA_k\}_{k=0}^{n-1} \in \C^{n_1 \times n_2}$ be an orthonormal basis of Hankel matrices \cite{ye2017compressive,cai2019fast}, which is defined as 
 \begin{align*}
 \boldsymbol{A}_{k}=\frac{1}{\sqrt{w_{k}}} \mathcal{H} (\ve_{k}),\, k \in \{0,\ldots,n-1\}
 \end{align*}
 where 
 \begin{equation*}
 w_{k}=|\{(i, j) | i+j=k,
 0 \leq i \leq n_{1}-1,0 \leq j \leq n_{2}-1\}|,
 \end{equation*}
 and $|S|$ returns the cardinality of the set $S$. Then $\mathcal{H}(\vx)$ can be expressed as
\begin{equation}
\mathcal{H}(\vx)= \sum_{k=0}^{n-1} \mathcal{A}_k(\mathcal{H}(\vx))= \sum_{k=0}^{n-1} \ip{\mathcal{H}(\vx)}{\vA_k} \vA_k,
\end{equation}
where $\mathcal{A}_k(\vX) \triangleq \ip{\vX}{\vA_k} \vA_k$ for $\vX \in \C^{n_1 \times n_2}$.


Let $\mathcal{H} (\vx) = \vU \vSigma \vV^H$ denote the compact singular value decomposition (SVD) of $\mathcal{H} (\vx)$ with $\vU \in \C^{n_1 \times r}$, $\vSigma \in \R^{r \times r}$ and $\vV \in \C^{r \times n_2}$. Let the subspace $\mathcal{\mathcal{T}}$ denote the support of  $\mathcal{H} (\vx)$  and $\mathcal{\mathcal{T}}^\perp$ be its orthogonal complement. Let $\sign(\widetilde{\vX})=\widetilde{\vU} \widetilde{\vV}^H$ denote the sign matrix of $\widetilde{\vX}$,
where $\widetilde{\vX} = \widetilde{\vU} \widetilde{\vSigma} \widetilde{\vV}^H$ denotes the compact SVD of $\widetilde{\vX}$. 

In order to analyses the matrix completion problem theoretically, we need to introduce the standard incoherence condition as follows
\begin{equation} \label{eq: incoherence condition}
\begin{aligned}
\max_{1 \leq i \leq n_1} \norm{\vU^H\ve_{i}}_2^2 &\le \frac{\mu r}{n_1},\\
\max_{1 \leq i \leq n_2} \norm{\vV^H \ve_{j}}_2^2 &\le \frac{\mu r}{n_2}.
\end{aligned}
\end{equation}
We also need to introduce the following norms which, respectively, measure the largest spectral norm among matrices $\{\mathcal{A}_k(\vX)\}_{k=1}^{n}$ and the $\ell_2$ norm of $\{\norm{\mathcal{A}_k(\vX)}\}_{k=1}^{n}$ \cite{chen2014robust,ye2017compressive}
\begin{equation}
\norm{\vX}_{\mathcal{A},\infty} \triangleq \max_{1 \leq k \leq n} \left| \ip{\vX}{\vA_k} \right| \norm{\vA_k},
\end{equation}
and
\begin{equation}
\norm{\vX}_{\mathcal{A},2} \triangleq \left( \sum_{k=1}^n \left| \ip{\vX}{\vA_k} \right|^2 \norm{\vA_k}^2 \right)^{1/2}.
\end{equation}


\section{Theoretical Guarantees} \label{sec: Performance guarantees}
In this section, we start by giving the theoretical guarantees for the proposed method. Then we extend the analysis to noisy circumstance. Our main result shows that when the prior information is reliable, the proposed approach \eqref{P: nuclear norm minimization with MC} can outperform previous approach \eqref{P: nuclear norm minimization} by a logarithmic factor. 

\begin{theorem}
	\label{thm:uniqueness}
	Let $\mathcal{H} (\vx)$ be a rank-$r$ matrix and satisfy the standard incoherence condition in Eq. \eqref{eq: incoherence condition} with parameter $\mu$. Consider a multi-set $\Omega = \{j_1,\ldots,j_m\}$  whose indicies $\{j_k\}_{i=1}^{m}$ are i.i.d. and follow the uniform distribution on $\{0,\ldots,n-1\}$.
	If the sample size satisfies
	\begin{multline*}
	m \ge \max\{\Delta^2,1\} \, c\mu c_{\mathrm{s}} r\log^3 n  \max\left\{\log \(7n \norm{\boldsymbol{F}_0}_F\),1\right\},
	\end{multline*}
	and the prior information satisfies
	$$\norm{\mathcal{P}_{\mathcal{T}^{\perp}}(\lambda \mathcal{G}(\vphi))}< \frac{1}{2},$$
	where $c>0$ is an absolute constant,
	$$c_s \triangleq \max\left\{\frac{n}{n_1}, \frac{n}{n_2}\right\},~ \boldsymbol{F}_{0}\triangleq \mathcal{P}_{\mathcal{T}}\left(\mbox{\upshape sgn}[\mathcal{H} (\vx)]-\lambda \mathcal{G}(\vphi) \right),$$
	and
	\begin{equation*}
	\Delta \triangleq \frac{4(\norm{ \boldsymbol{F}_{0}}_{\mathcal{A},2}+\norm{ \boldsymbol{F}_{0}}_{\mathcal{A},\infty})}{1-2 \norm{\mathcal{P}_{\mathcal{T}^{\perp}}(\lambda \mathcal{G}(\vphi))}},
	\end{equation*}	
	then $\vx$ is the unique minimizer for the approach \eqref{P: nuclear norm minimization with MC} with high probability.
	
\end{theorem}

\begin{remark}[{Comparison with \cite[Theorem 1]{ye2017compressive}}] When there is no prior information or no reliable prior information, $\vphi$ is set to be $\bm{0}$ and the program \eqref{P: nuclear norm minimization with MC} would reduce to \eqref{P: nuclear norm minimization}. However, when the prior information is reliable, the proposed approach can reduce the sampling size by $O(\log n)$ compared with the results \cite[Theorem 1]{ye2017compressive}.
\end{remark}

\begin{remark}[The choice of operator $\mathcal{F}$] \label{rm:operator}  It should be noted that the choice of operator $\mathcal{F}$ will influence the performance of the proposed program. According to the definition of $\vF_0$, it's not hard to see that $\mathcal{F}(\cdot)=\sign(\cdot)$ is a suitable choice to improve the sampling bound. In this case, the value of $\norm{\boldsymbol{F}_0}_F$ will be very small when the subspace information of ${\mathcal{H}(\vphi)}$ is very similar to that of ${\mathcal{H}(\vx)}$ and $\lambda=1$.
Accordingly, the program becomes  	
\begin{equation*}
\begin{aligned} 
&\min \limits_{\vz} ~~ \norm{\mathcal{H}(\vz)}_*-2 \lambda \, \text{Re}\(\ip{\sign({\mathcal{H}(\vphi)})}{\mathcal{H}(\vz)}\) \\
&~\text{s.t.}~~~  \mathcal{P}_\Omega(\vz)= \mathcal{P}_\Omega({\vx}).
\end{aligned}
\end{equation*}	
\end{remark}

\begin{remark}[The choice of weight $\lambda$]  Note that the sampling lower bound is determined by the value of $\norm{\vF_0}_F$ and the best choice of $\lambda$ is the one that minimizes $\norm{\vF_0}_F$. The expression of $\norm{\vF_0}_F^2$ can be rewritten as 
\begin{multline*}
\|\vF_0\|_F^2=\lambda^2 \norm{\mathcal{P}_{\mathcal{T}}(\mathcal{G}(\vphi))}_F^2+\norm{\mathcal{P}_{\mathcal{T}}(\mbox{\upshape sgn}[\mathcal{H} (\vx)])}_F^2 \\-2\lambda\, \mbox{Re}(\ip{\mathcal{P}_{\mathcal{T}}(\mbox{\upshape sgn}[\mathcal{H} (\vx)])}{\mathcal{G}(\vphi)}).
\end{multline*}
So the optimal weight is
\begin{equation*}
	\lambda^\star =\frac{\mbox{Re}\(\ip{\mathcal{P}_{\mathcal{T}}(\mbox{\upshape sgn}[\mathcal{H} (\vx)])}{\mathcal{G}(\vphi)}\)}{\norm{\mathcal{P}_{\mathcal{T}}(\mathcal{G}(\vphi))}_F^2}.
\end{equation*}
Let $\mathcal{G}(\vphi)=\sign({\mathcal{H}(\vphi)})$ as Remark \ref{rm:operator}. When the prior information is close to the desired signal, $\lambda$ should be around $1$. On the contrary, when the prior information is extremely different from the desired signal, $\lambda$ should be around $0$.

\end{remark}

\begin{remark}[The wrap-around operator] When $\mathcal{H}\(\cdot\)$ is replaced with the following operator $\mathcal{H}_c (\cdot)$ with the wrap-around property
	$$\mathcal{H}_c (\vx) \triangleq \left[ \begin{array}{cccc}
	{x_0} & {x_1} & {\ldots} & {x_{d-1}}  \\ 
	{x_1} & {x_2} & {\ldots} & {x_d}  \\
	{\vdots} & {\vdots} & {\ddots} & {\vdots}  \\
	{x_{n-d}} & {x_{n-d+1}} & {\ldots} & {x_{n-1}}  \\	
	{x_{n-d+1}} & {x_{n-d+2}} & {\ldots} & {x_{0}}  \\
	{\vdots} & {\vdots} & {\ddots} & {\vdots}  \\
	{x_{n-1}} & {x_{0}} & {\ldots} & {x_{d-2}}  \\			
	\end{array} \right],$$
	where $\mathcal{H}_c (\vx) \in \C^{n\times d}$, it is straightforward to obtain the lower bound for sample size by following the proof in \cite{ye2017compressive}
	\begin{multline*}
	m \ge \max\{\Delta^2,1\} \, c \mu c_{\mathrm{s}} r\log n  \max\left\{\log \(7n \norm{\boldsymbol{F}_0}_F\),1\right\}.
	\end{multline*}
	In this case, $ O(r \log n)$ samples are enough to exactly reconstruct the original signals when the prior information is reliable, which outperforms the atomic norm minimization in \cite{tang2013compressed,tang2014near}.
	
\end{remark}

\begin{table*}[!htb]
	\begin{align}\label{eq: Mult-level_Hankel} \mathcal{H}^{d}\left(\mathcal{X}^{d}\right)= \left[\begin{array}{cccc}
	{\mathcal{H}^{d-1}\left(\mathcal{X}^{d-1}(0)\right)} & {\mathcal{H}^{d-1}\left(\mathcal{X}^{d-1}(1)\right)} & 
	{\cdots} & 
	{\mathcal{H}^{d-1}\left(\mathcal{X}^{d-1}\left(n_d-1\right)\right)} \\ {\mathcal{H}^{d-1}\left(\mathcal{X}^{d-1}(1)\right)} & 
	{\mathcal{H}^{d-1}\left(\mathcal{X}^{d-1}(2)\right)} &
	{\cdots} & 
	{\mathcal{H}^{d-1}\left(\mathcal{X}^{d-1}\left(n_d\right)\right)} \\ 
	{\vdots} & {\vdots} &{\ddots} & {\vdots} \\ {\mathcal{H}^{d-1}\left(\mathcal{X}^{d-1}\left(N_{d}-n_d\right)\right)} &
	{\mathcal{H}^{d-1}\left(\mathcal{X}^{d-1}\left(N_{d}-n_d+1\right)\right)} &
	{\cdots} & 
	{\mathcal{H}^{d-1}\left(\mathcal{X}^{d-1}(N_d-1)\right)}\end{array}\right] 
	\end{align}
	\hrulefill
\end{table*}

By straightly following \cite[Theorem 7]{candes2010matrix}, an extension to the noisy version  with bounded noise can be shown as follows.

\begin{corollary}
\label{thm:robust recovery}
Let $\mathcal{H} (\vx)$ be a rank-$r$ matrix and satisfy the standard incoherence condition in Eq. \eqref{eq: incoherence condition} with parameter $\mu$. Consider a multi-set $\Omega = \{j_1,\ldots,j_m\}$  whose indicies $\{j_k\}_{i=1}^{m}$ are i.i.d. and follow the uniform distribution on $\{0,\ldots,n-1\}$.  Suppose the noisy observation $\vy=\vx+\vn$, where $\vn$ denotes bounded noise. Let $\vx^\dagger$ be the solution of the noisy version program
\begin{equation} \label{P: noisy nuclear norm minimization with MC}
\begin{aligned} 
&\min \limits_{\vz} ~~ \norm{\mathcal{H}(\vz)}_*-2 \lambda \, \text{\rm Re}\(\ip{\mathcal{G}(\vphi)}{\mathcal{H}(\vz)}\) \nonumber\\
&~\text{\rm s.t.}~~~  \norm{\mathcal{P}_\Omega(\vz)- \mathcal{P}_\Omega({\vy})}_2 \le \delta.
\end{aligned}
\end{equation}
If the sample size satisfies
\begin{multline*}
m \ge \max\{\Delta^2,1\} \, \mu c_{\mathrm{s}} r\log^3 n  \max\left\{\log \(7n \norm{\boldsymbol{F}_0}_F\),1\right\},
\end{multline*}
and the prior information satisfies
$$\norm{\mathcal{P}_{\mathcal{T}^{\perp}}(\lambda \mathcal{G}(\vphi))}< \frac{1}{2},$$	
then the solution $\vx^\dagger$ satisfies that
$$
\norm{\mathcal{H}(\vx)-\mathcal{H}(\vx^\dagger)}_F \le c \delta \, \(n^2+n^{\frac{3}{2}} \norm{\lambda \mathcal{G}(\vphi)}_F\)
$$
with high probability.
\end{corollary}

\section{Extensions to multi-dimensional models} \label{sec: multi-dimensional models}
In this section, we extend the analysis from one-dimensional signal to multi-dimensional signal. Consider a $d$-way tensor $\mathcal{X}^d \in \C^{N_1\times \ldots\times N_d}$, each of whose entries can be denoted as
$$
\mathcal{X}^d(k_1,\ldots,k_d)=\sum_{l=1}^{r}w_l e^{j2\pi  \sum_{j=1}^{d} k_j f_{j,l}},
$$
for $(k_1,\ldots,k_d) \in \{1,\ldots,N_1\}\times \ldots \times\{1,\ldots,N_d\}$. Denote $\vf_l=(f_{1,l},\ldots,f_{d,l}) \in [0,1)^d$ as the frequency vector for $l=1,\ldots,r$.

Similar to the one-dimensional case, we use multi-level Hankel operator to lift $\mathcal{X}^d$ to a low-rank matrix. See Eq. \eqref{eq: Mult-level_Hankel},
where $\mathcal{X}^{d-1}(i)=\mathcal{X}^{d}(:,\ldots,:,i),\,i=0,\ldots,\max\{n_d-1,N_d-n_d\}$. When $d=1$, the above operator degrades to normal Hankel operator. According to \cite{chen2014robust}, the rank of the lifted matrix satisfies $\mbox{Rank}(\mathcal{H}^{d}\left(\mathcal{X}^{d}\right)) \le r$ by using high-dimensional Vandermonde decomposition. Let $\Phi^d$ denote the prior information of $\mathcal{X}^d$. We can complete the $d$-way tensor by using the following nuclear norm minimization
\begin{equation}\label{P: nuclear norm minimization with MC MD}
\begin{aligned} 
&\min \limits_{\mathcal{Z}^d} ~~ \norm{\mathcal{H}^{d}(\mathcal{Z}^d)}_*-2\lambda \, \text{Re}\(\ip{{\mathcal{G}(\Phi^d)}}{\mathcal{H}^{d}(\mathcal{Z}^d)}\) \\
&~\text{s.t.}~~~  \mathcal{P}_\Omega(\mathcal{Z}^d)= \mathcal{P}_\Omega(\mathcal{X}^d),
\end{aligned}
\end{equation}
where $\Omega=\{(k_1,\ldots,k_d) \in \{0,\ldots,N_1-1\}\times \ldots \times\{0,\ldots,N_d-1\}\}$ denotes the index set of known entries of $\mathcal{X}^d$ and $\ip{\mathcal{X}^d}{\mathcal{Y}^d}=(\mbox{vec} (\mathcal{Y}^d))^H \mbox{vec} (\mathcal{X}^d)$ denotes the inner product of the $d$-way tensors.

Let $\bm{E}(k_1,\ldots,k_d)$ be the canonical basis in the domain $\C^{N_1\times\ldots\times N_d}$, and define the following orthonormal basis,
$$
\vA(k_1,\ldots,k_d)=\frac{1}{\sqrt{w(k_1,\ldots,k_d)}} \mathcal{H}(\bm{E}(k_1,\ldots,k_d)),
$$
where 
\begin{align*}
w(k_1,\ldots,k_d)= \prod_{l=1}^{d} w_l
\end{align*}
and
\begin{multline*}
w_{l}=| \{(i_l, j_l) | i_l+j_l=k_l,
0 \leq i_l \leq n_{l}-1,\\
0 \leq j_l \leq N_{l}-n_l+1\}|.
\end{multline*}

So each $d$-way tensor $\mathcal{X}^d$ can be rewritten as
\begin{multline*}
\mathcal{H}(\mathcal{X}^d)\\
=\sum_{k_1=0}^{N_1-1}\ldots\sum_{k_d=0}^{N_d-1}  \ip {\mathcal{H}(\mathcal{X}^d)}{\vA(k_1,\ldots,k_d)} \vA(k_1,\ldots,k_d).
\end{multline*}

It's straightforward to extend the theoretical guarantee from one-dimensional case to multi-dimensional case.
\begin{theorem}
	\label{thm:uniqueness0fMD}
	Let $\Omega = \{\vj_1,\ldots,\vj_m\}$ be a multi-set consisting of random indices
	where $\{\vj_k\}_{k=1}^{m} \in \R^d$ are i.i.d. and follow the uniform distribution on $[N_1]\times\ldots\times [N_d]$.
	Suppose, furthermore, that  $\mathcal{H} (\vx)$ is of rank-$r$ and satisfies the standard incoherence condition in \eqref{eq: incoherence condition} with parameter $\mu$.
	Then there exists an absolute constant $c_1$ such that
	$\vx$ is the unique minimizer to \eqref{P: nuclear norm minimization with MC} with high probability, 
	provided that
	\begin{multline*}
	m \ge \max\{\Delta^2,1\} \, c \mu c_{\mathrm{s}} r\log^\alpha \(\prod_{k=1}^d N_k\) \\ \cdot \max\left\{\log \(7\prod_{k=1}^d N_k \norm{\boldsymbol{F}_0}_F\),1\right\},
	\end{multline*}
	and
	$$\norm{\mathcal{P}_{\mathcal{T}^{\perp}}(\lambda \mathcal{G}(\vphi))}< \frac{1}{2},$$
	where
	\begin{align*}
	c_s &\triangleq \max\left\{\prod_{k=1}^d \frac{N_k}{n_k} , \prod_{k=1}^d \frac{N_k}{N_k-n_k+1}\right\},\\
	\boldsymbol{F}_{0} &\triangleq \mathcal{P}_{\mathcal{T}}\left(\mbox{\upshape sgn}[\mathcal{H} (\vx^\star)]-\lambda \mathcal{G}(\vphi) \right),
	\end{align*}
	and
	\begin{equation*}
	\Delta \triangleq \frac{4(\norm{ \boldsymbol{F}_{0}}_{\mathcal{A},2}+\norm{ \boldsymbol{F}_{0}}_{\mathcal{A},\infty})}{1-2 \norm{\mathcal{P}_{\mathcal{T}^{\perp}}(\lambda \mathcal{G}(\vphi))}}.
	\end{equation*}
	Here, $\alpha = 1$ if the lifting operator has the wrap-around property; $\alpha = 3$ if the lifting operator doesn't have the wrap-around property.	
\end{theorem}

\section{Optimization Algorithm} \label{sec: Optimization Algorithm}
Due to the high computational complexity of low rank methods, we decide to use the non-convex method to solve the problem. First we use matrix factorization to decompose $\mathcal{H}(\vz)$ to two low complexity matrices, i.e.  $\mathcal{H}(\vz)=\vU\vV^H$ with $\vU \in \C^{n_1 \times r}$ and $\vV \in \C^{n_2 \times r}$. Then we use  Alternating Direction Method of Multipliers (ADMM) \cite{boyd2011distributed} to solve the problem.

First of all, we denote the nuclear norm as follows \cite[Lemma 8]{srebro2004learning}
\begin{equation} \label{eq: nuclear norm decomposition}
	\norm{\mathcal{H}(\vz)}_*=\min_{\vU,\vV:\mathcal{H}(\vz)=\vU\vV^H} \frac{1}{2}(\norm{\vU}^2_F+\norm{\vV}^2_F).
\end{equation}

Incorporating \eqref{eq: nuclear norm decomposition} into the problem \eqref{P: nuclear norm minimization with MC} yields
\begin{align*}
&\min \limits_{\vz,\vU,\vV} ~~ \frac{1}{2}(\norm{\vU}^2_F+\norm{\vV}^2_F)-\lambda \, \text{Re}\(\ip{\mathcal{G}(\vphi)}{\vU \vV^H}\) \nonumber\\
&~~\text{s.t.}~~~  \mathcal{P}_\Omega(\vz)= \mathcal{P}_\Omega({\vx}),~ \mathcal{H}(\vz)=\vU\vV^H.
\end{align*}

Then, we start ADMM by the following argumented Lagrange function
\begin{multline} \label{eq: argumented Lagrange function}
L(\vU,\vV,\vz,\vLambda)= \Pi(\vz)-2 \lambda \, \text{Re}\(\ip{\mathcal{G}(\vphi)}{\vU \vV^H}\) 
 \\+\frac{1}{2}(\norm{\vU}^2_F+\norm{\vV}^2_F) + \frac{\mu}{2}\norm{\mathcal{H}(\vz)-\vU\vV^H+\vLambda}_F^2,
\end{multline}
where $\mu>0$ is an absolute constant, and $\Pi(\vz)$ is an indicator function
\begin{equation*}
\Pi(\vz)=
\left\{
{\begin{array}{rl}
	0, & \text{if} ~\mathcal{P}_\Omega(\vz)= \mathcal{P}_\Omega({\vx}),  \\
	\infty, & \text{otherwise}.  \\
	\end{array} }
\right.
\end{equation*}

Next, we decompose \eqref{eq: argumented Lagrange function} into three subproblems to get  $\vz^{(n+1)}$, $\vU^{(n+1)}$ and $\vV^{(n+1)}$
\begin{align*}
	\vz^{(n+1)}&= \mathop{\arg\min}_{\vz} \Pi(\vz) + \frac{\mu}{2}\norm{\mathcal{H}(\vz)-\vU^{(n)}\vV^{(n)H}+\vLambda^{(n)}}_F^2\\
	\vU^{(n+1)}&= \mathop{\arg\min}_{\vU} \frac{1}{2}\norm{\vU}^2_F -2 \lambda \, \text{Re}\(\ip{\mathcal{G}(\vphi)}{\vU \vV^{(n)H}}\) \\
	&\quad + \frac{\mu}{2} \norm{\mathcal{H}(\vz^{(n+1)})-\vU\vV^{(n)H}+\vLambda^{(n)}}_F^2\\
	\vV^{(n+1)}&= \mathop{\arg\min}_{\vV} \frac{1}{2}\norm{\vV}^2_F-2 \lambda \, \text{Re}\(\ip{\mathcal{G}(\vphi)}{\vU^{(n+1)} \vV^H}\) \\
	&\quad + \frac{\mu}{2} \norm{\mathcal{H}(\vz^{(n+1)})-\vU^{(n+1)}\vV^H+\vLambda^{(n)}}_F^2
\end{align*}
and the Lagrangian update is 
\begin{equation*}
	\vLambda^{(n+1)}=\mathcal{H}(\vz^{(n+1)})-\vU^{(n+1)}\vV^{(n+1)H}+\vLambda^{(n)}.
\end{equation*}

\begin{algorithm}[t]
	\caption{Reference-based Structured Matrix Completion}
	\label{alg: I}
	\begin{algorithmic}[1]
		\REQUIRE  Sampling index set $\Omega$, measurements  $\mathcal{P}_\Omega({\vx})$, prior information $\vphi$
		\ENSURE   Estimated result $\vz$
		\STATE
		Initialize
		$k=0,$ $\varepsilon,$ $tol,$ and $K$
		\IF {$\norm{\mathcal{P}_\Omega(\vx-\vphi)}\le \varepsilon$}
		\STATE
		$[\vU_{(r)},\vSigma_{(r)},\vV_{(r)}]=r\text{-SVD}(\mathcal{H}(\vphi));$
		$\vLambda^{(0)}=\bm{0};$ $\vU_0=\vU_{(r)} \vSigma_{(r)};$ $\vV_0=\vV_{(r)};$ $G(\vphi)=\vU_{(r)}*\vV_{(r)}^H$
		\ELSE
		\STATE
		$[\vU_0,\vV_0]=\text{LMaFit}\(\mathcal{H},\mathcal{P}_\Omega({\vx})\);$
		$\vLambda^{(0)}=\bm{0};$ $G(\vphi)=\bm{0}$
		\ENDIF
		\REPEAT
		\STATE $k=k+1$
		\STATE $\vz_k= \mathcal{P}_{\Omega^c} \mathcal{H}^{\dagger}(\vU_{k-1}\vV_{k-1}^H-\vLambda_{k-1}) + \mathcal{P}_\Omega({\vx})$
		\STATE $\vU_k= \[\mu\(\mathcal{H}(\vz_k)+\vLambda_{k-1}\)+\lambda\mathcal{G}(\vphi)\]
		\cdot \vV_{k-1} \cdot(\vI+\mu \vV_{k-1}^{H}\vV_{k-1})^{-1} $
		\STATE $\vV_k= \[\mu\(\mathcal{H}(\vz_k)+\vLambda_{k-1}\)+\lambda\mathcal{G}(\vphi)\]^H \cdot \vU_k \cdot (\vI+\mu \vU_k^H \vU_k)^{-1}$
		\STATE $\vLambda_k=\mathcal{H}(\vz_k)-\vU_k\vV_k^{H}+\vLambda_{k-1}$	
		\UNTIL{$k>K$ or $\norm{\vz_k-\vz_{k-1}}_F<tol$}
	\end{algorithmic}
\end{algorithm}

By simple calculations, we can obtain
\begin{equation*}
	\vz^{(n+1)}= \mathcal{P}_{\Omega^c} \mathcal{H}^{\dagger}(\vU^{(n)}\vV^{(n)H}-\vLambda^{(n)}) + \mathcal{P}_\Omega({\vx}),
\end{equation*}
where $\mathcal{P}_{\Omega^c}$  denotes the projection operator on $\Omega^c$ and $\mathcal{H}^{\dagger}(\cdot)$ denotes the Penrose-Moore pseudo-inverse mapping corresponding to $\mathcal{H}(\cdot)$.

Then, by taking the derivative of the other two problems and setting them to zero, we can otain 
\begin{align*}
\vU^{(n+1)}&= \[\mu\(\mathcal{H}(\vz^{(n+1)})+\vLambda^{(n)}\)+\lambda\mathcal{G}(\vphi)\]\\
& \qquad \qquad \qquad \quad \cdot \vV^{(n)} (\vI+\mu \vV^{(n)H}\vV^{(n)})^{-1}
\end{align*}
and 
\begin{align*}
\vV^{(n+1)}&= \[\mu\(\mathcal{H}(\vz^{(n+1)})+\vLambda^{(n)}\)+\lambda\mathcal{G}(\vphi)\]^H\\
& \qquad \qquad \cdot \vU^{(n+1)} (\vI+\mu \vU^{(n+1)H}\vU^{(n+1)})^{-1}.
\end{align*}

The last question is how to initialize $\vU$ and $\vV$. In order to converge quickly, the authors in \cite{Jin2016General} uses an algorithm named LMaFit \cite{wen2012solving}, which is
\begin{equation} \label{LMaFit}
\min_{\vU,\vV,\vZ} \frac{1}{2} \norm{\vU\vV^H-\vZ}_F^2~~ \text{s.t.}~\mathcal{P}_\Omega(\mathcal{H}^{\dagger}(\vZ)) = \mathcal{P}_\Omega(\vx).
\end{equation}

However, LMaFit only uses the undersampled measurements and cannot guarantee that $\vZ$ has the lifting matrix structure. Instead, we can take advantage of the reference image to initialize $\vU$ and $\vV$ by using truncated SVD when the reference image is reliable. Here, we use the value of $\norm{\mathcal{P}_\Omega(\vx-\vphi)}$ as a criterion to choose the suitable initialization strategy.

Then we can give the corresponding algorithm in Algorithm \ref{alg: I}. Here, $r\text{-SVD}(\cdot)$ returns the results of truncated SVD.  And $\text{LMaFit}(\mathcal{H},\mathcal{P}_\Omega({\vx}))$ denotes the algorithm in \eqref{LMaFit}.

\begin{figure*}[!t]
	\centering
	\subfloat[]{
		\includegraphics[width=2.2in]{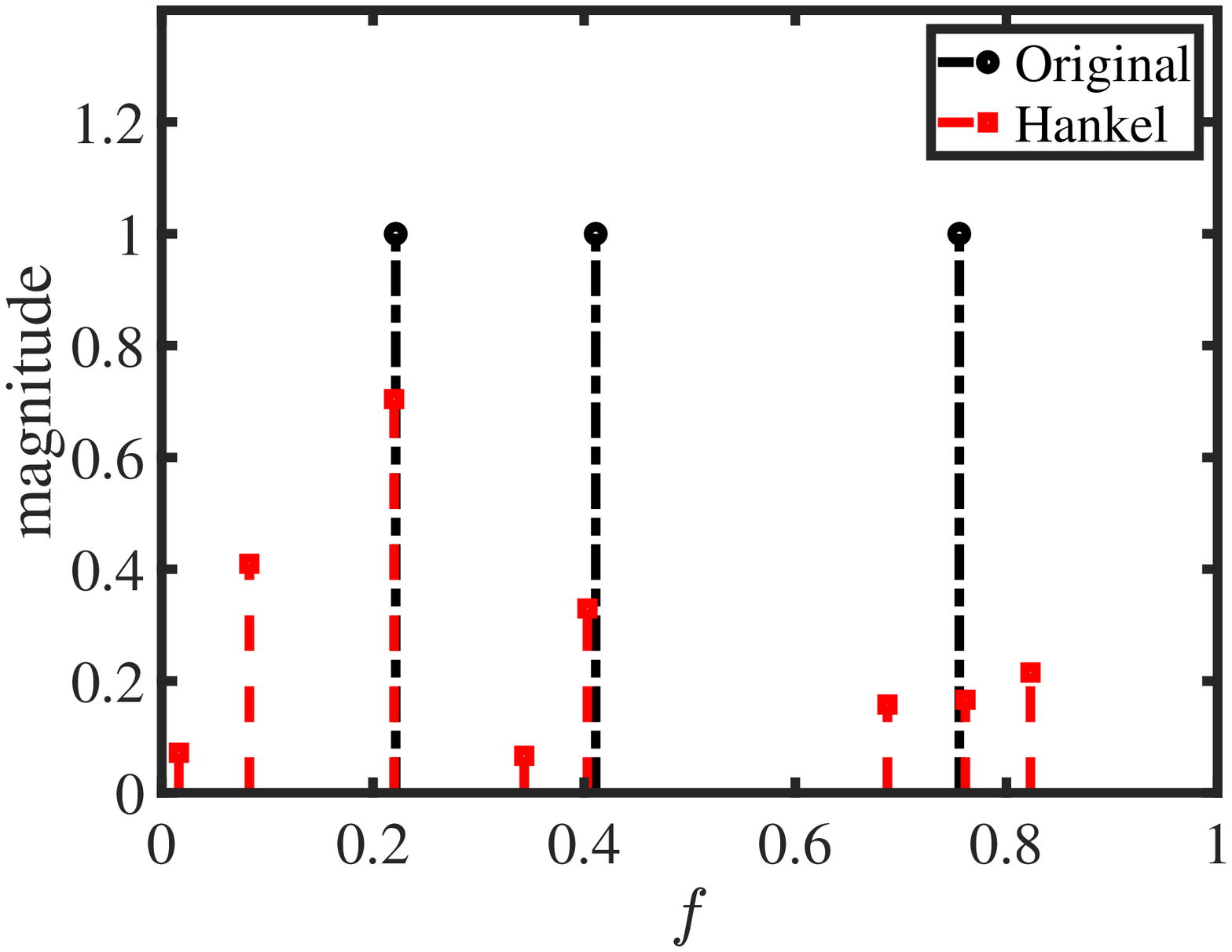}}
	\hfil
	\subfloat[]{
		\includegraphics[width=2.2in]{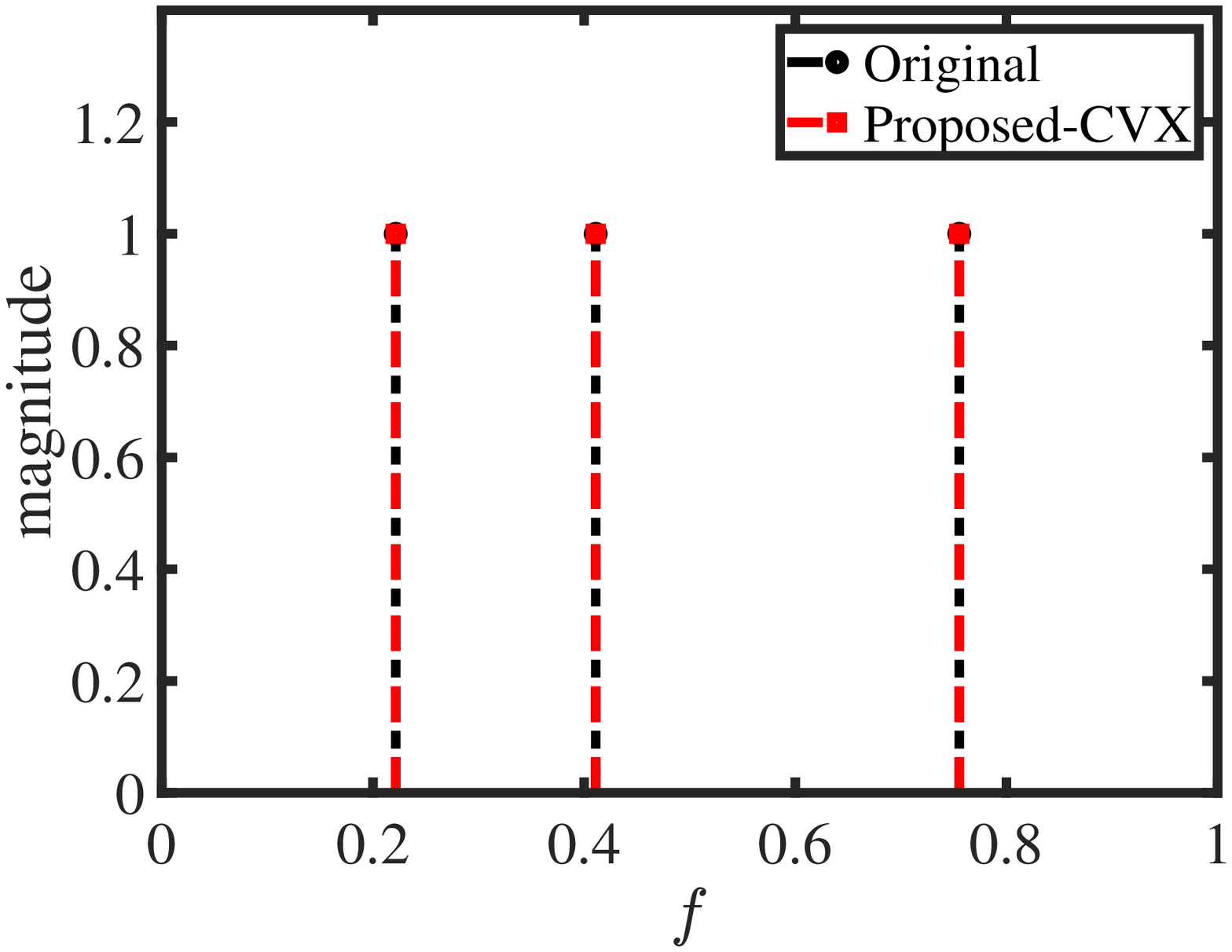}}
	\hfil
	\subfloat[]{
		\includegraphics[width=2.2in]{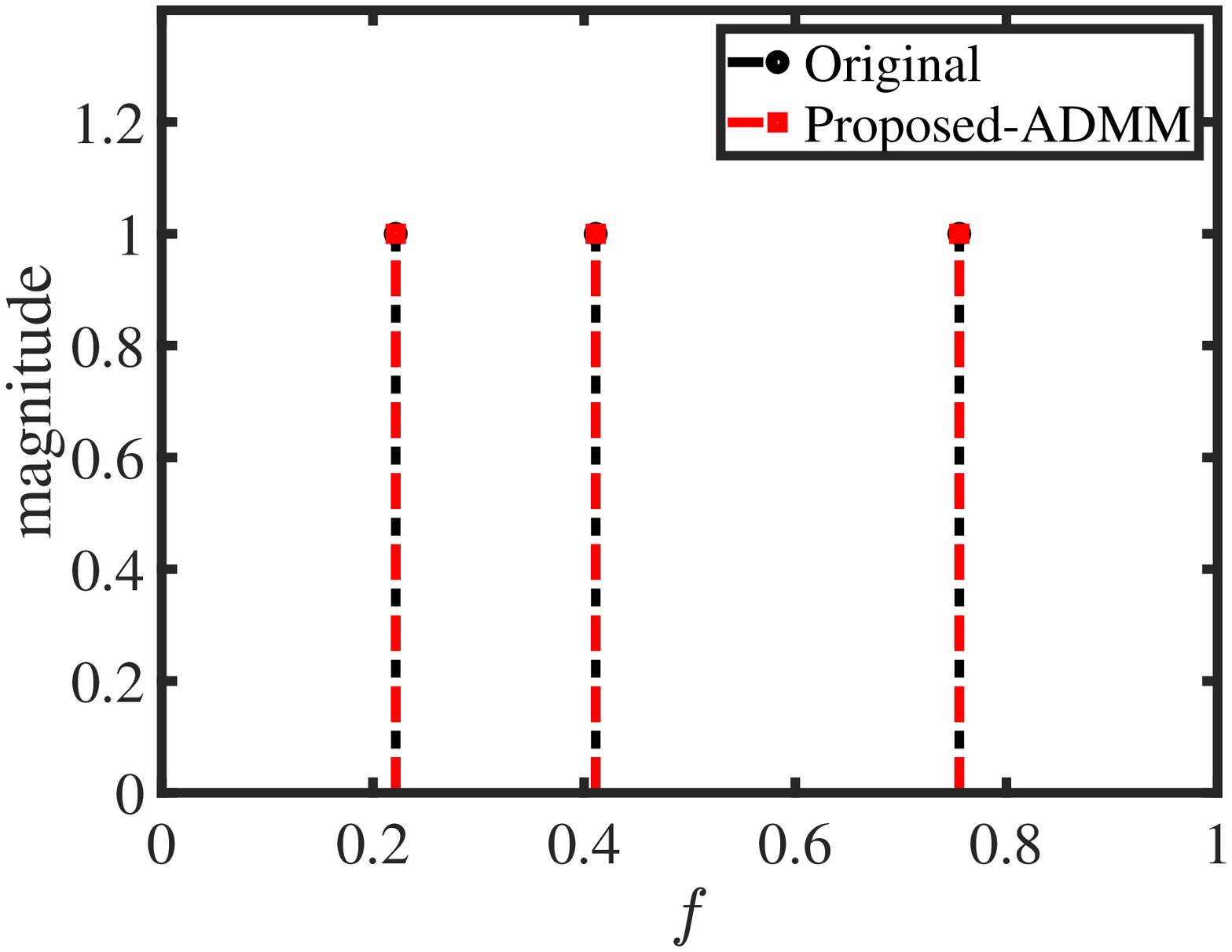}}
	\caption{Performance Comparisons for one-dimensional signals when $n=32$, $r=3$ and $m=10$. (a) Hankel matrix completion; (b) Proposed method (CVX); (c) Proposed method (ADMM).}
	\label{fig: PerformanceComparison_001}
\end{figure*}

\begin{figure}[!t]
	\centering
	\includegraphics[width=2.5in]{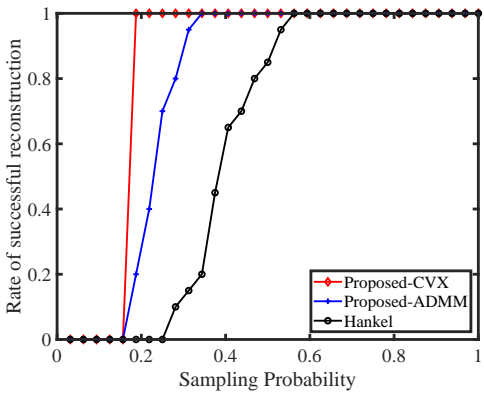}
	\caption{Rate of successful reconstruction v.s. sampling probability  for Hankel matrix completion and reference based Hankel matrix completion. }
	\label{fig: PerformanceComparison_002}
\end{figure}

\begin{figure*}[!t]
	\centering
	\subfloat[]{
		\includegraphics[width=2.3in]{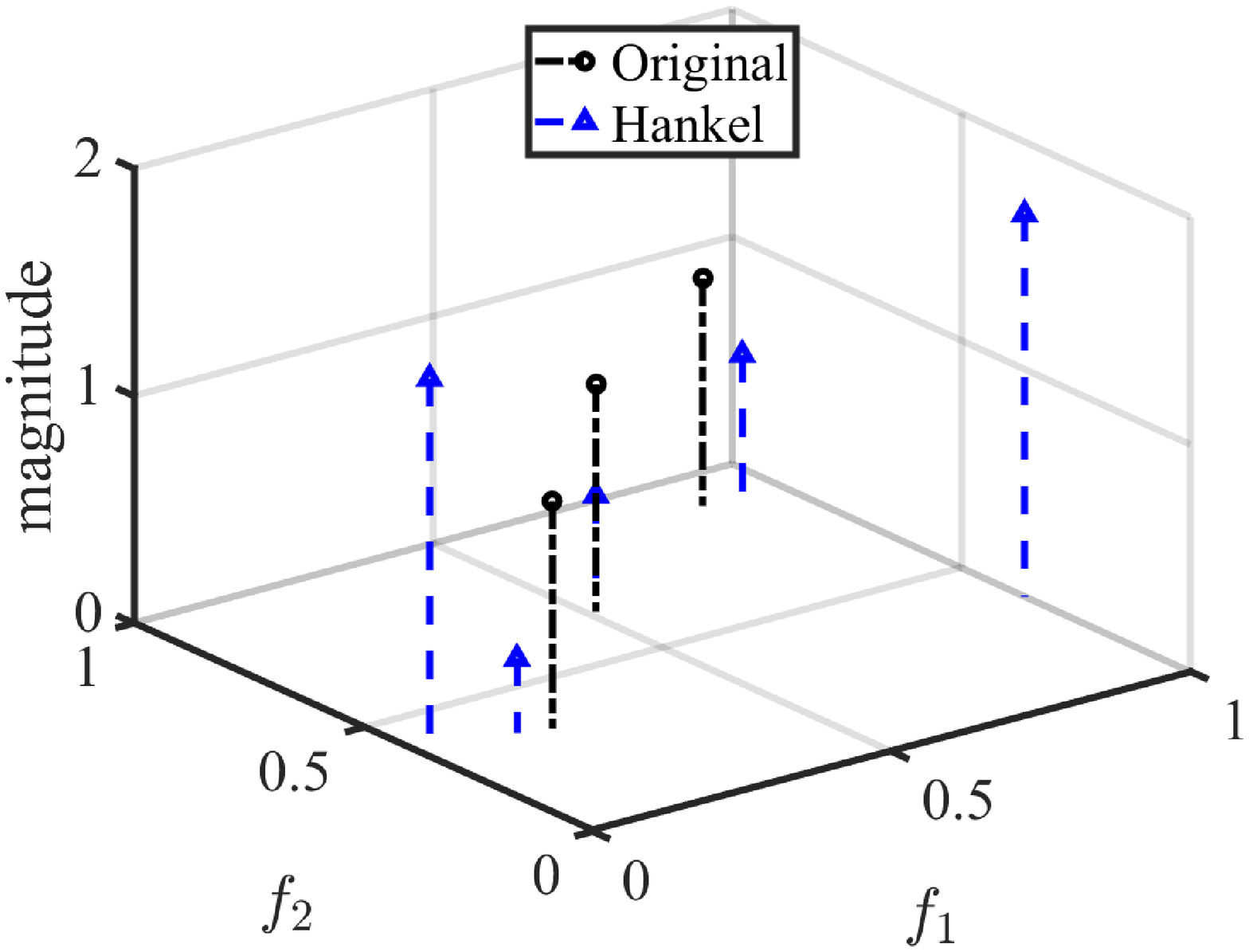}}
	\hfil
	\subfloat[]{
		\includegraphics[width=2.3in]{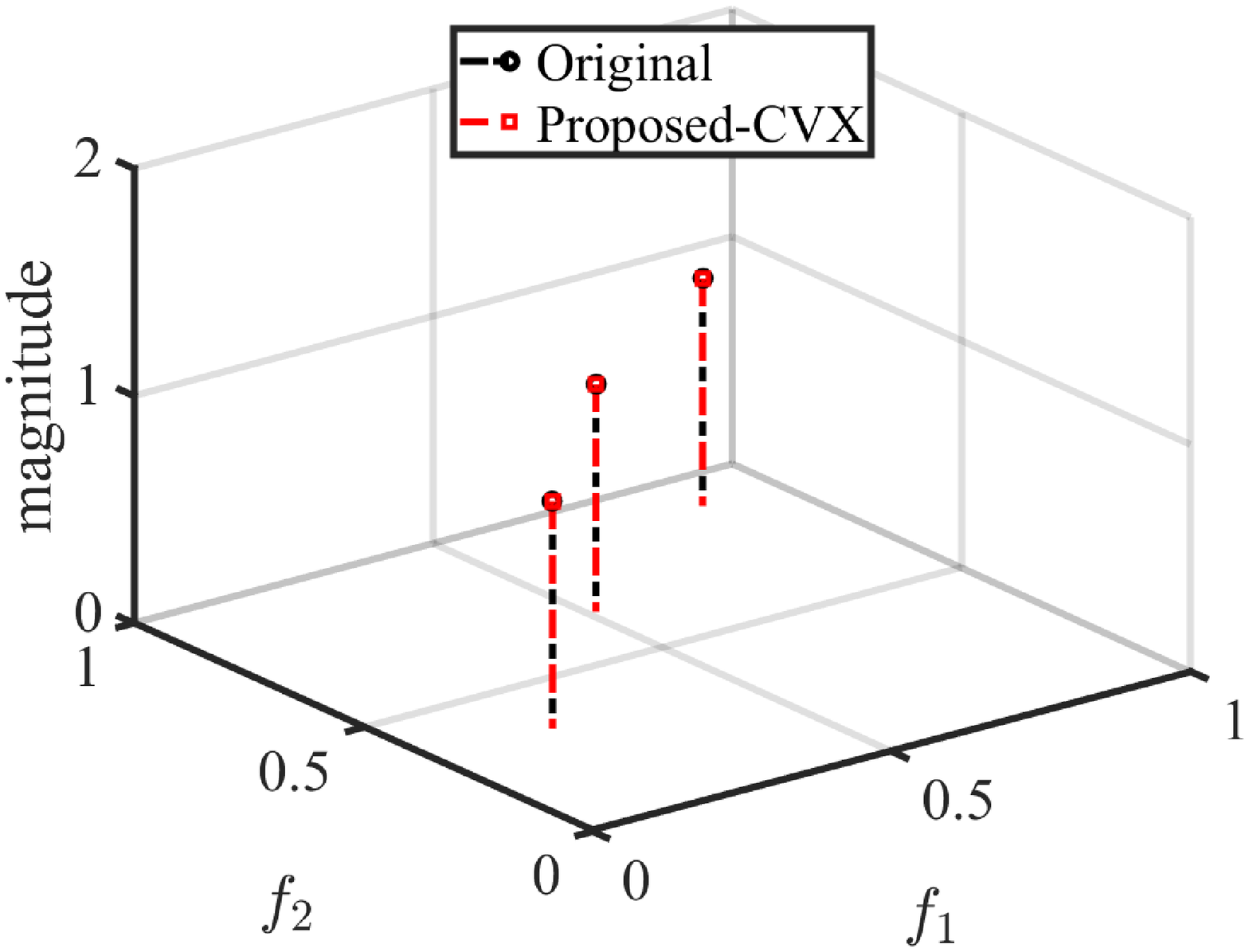}}
	\hfil
	\subfloat[]{
		\includegraphics[width=2.3in]{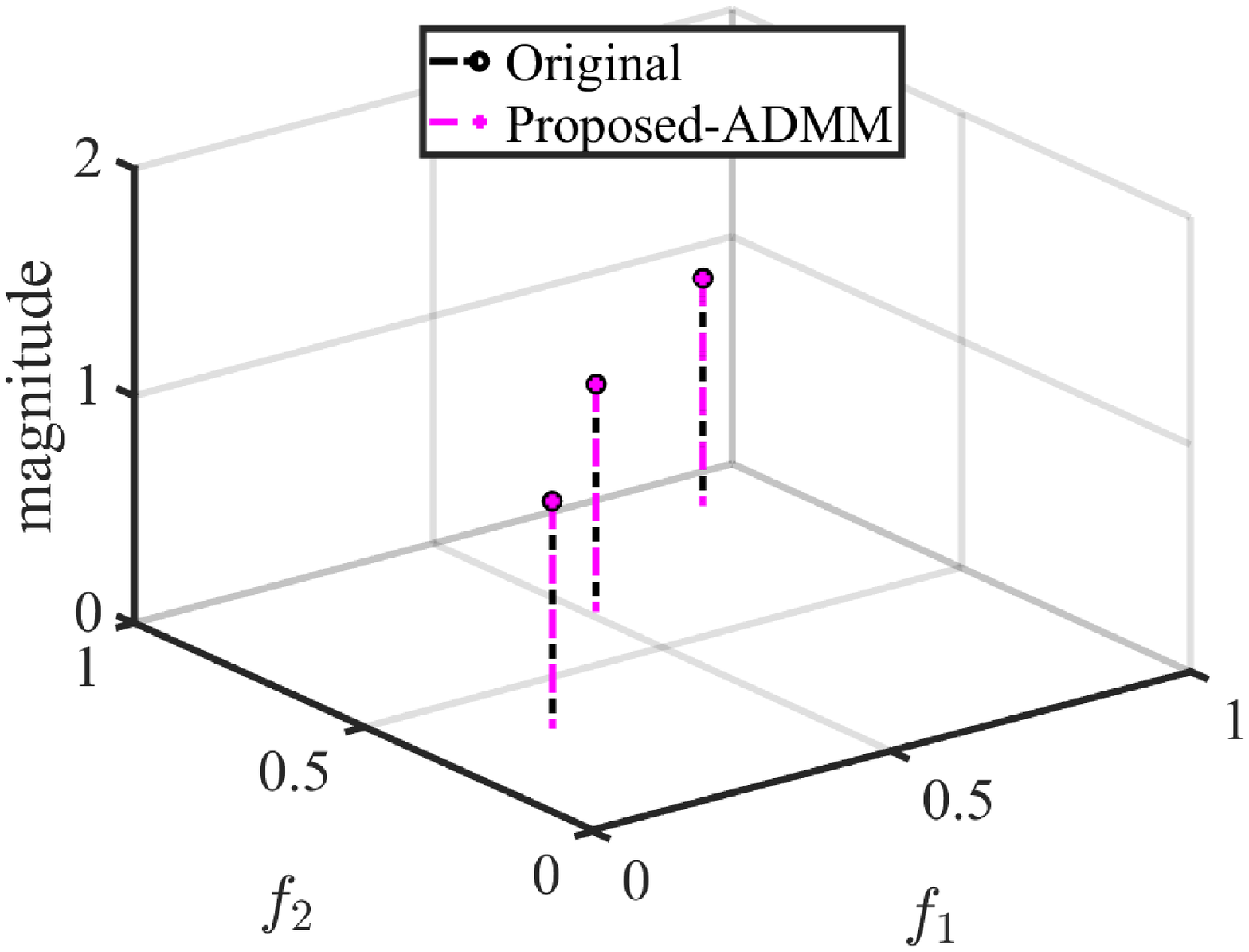}}
	\caption{Performance Comparisons for two-dimensional signals when $N_1=10, N_2=10, r=3, m=20$. (a) Hankel matrix completion; (b) Proposed method (CVX); (c) Proposed method (ADMM).}
	\label{fig: PerformanceComparison_003}
\end{figure*}

\begin{figure}[!t]
	\centering
	\includegraphics[width=2.5in]{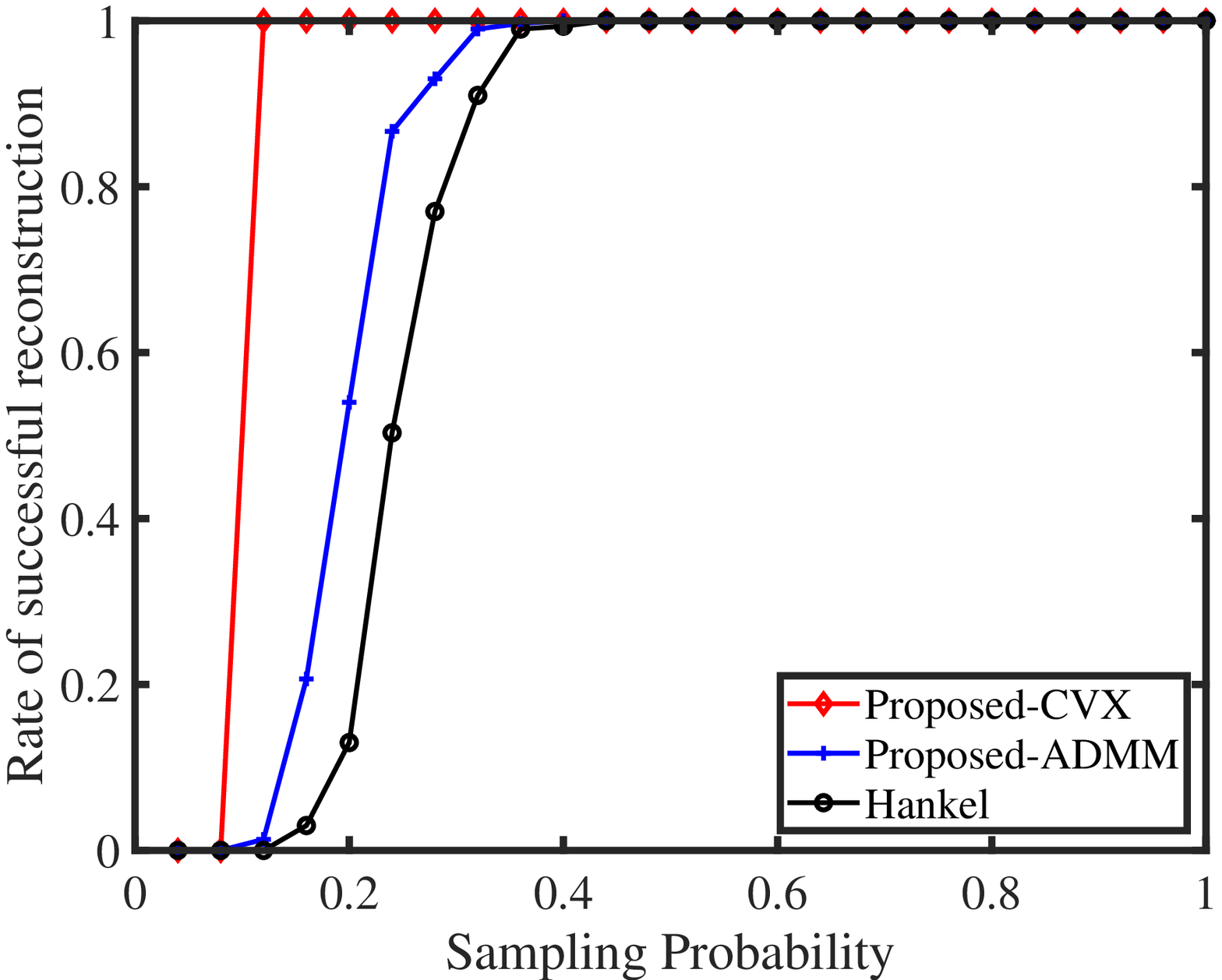}
	\caption{Phase transitions for Hankel matrix completion and reference based Hankel matrix completion. }
	\label{fig: PerformanceComparison_004}
\end{figure}

\section{Simulations} \label{sec: Simulation}
In this section, we carry on numerical simulations to show the improvement of the proposed method \eqref{P: nuclear norm minimization with MC} compared to standard Hankel matrix completion \eqref{P: nuclear norm minimization}. Besides, we compare the performance under two different solvers: CVX solver and ADMM-solver. Here, we use CVX package\cite{cvx,gb08} to get the convex results and use Algorithm 1 to get the ADMM results.

\subsection{Simulations for 1-D signals}
We begin by giving the numerical results for one-dimensional signals.  

Consider a one-dimensional spectrally sparse signal $\vx^\star \in \C^n$ and the signal is a weighted superposition of $r$ complex sinusoids with unit amplitudes. The reference signal is created by $\vphi=\vx^\star+\sigma \vn \in \C^n$, where the entries of the real and imaginary part of $\vn$ follow i.i.d. standard Gaussian distribution, i.e., $ \mbox{Re}(n_i), \mbox{Im}(n_i) \sim \mathcal{N}(0,1)$ for $i=1,\ldots,n$. 

We first show the reconstruction results for standard Hankel matrix completion, the proposed method with CVX solver and the proposed method with ADMM solver.  We set $n=32$, $r=3$, $m=10$ and $\sigma=0.5$. The  matrix pencil method is used to estimate the location and amplitude of frequencies \cite{hua1990matrix}. The frequency estimation results are shown in Fig. \ref{fig: PerformanceComparison_001}. As expected, with the reliable reference signal, the proposed scheme with different solvers exactly reconstructs the original signal,which has a better performance than standard Hankel matrix completion.

We next provide the successful reconstruction rate as a function of sampling probability standard Hankel matrix completion, the proposed method with CVX solver and the proposed method with ADMM solver.   We set $n=32$, $r=3$, $\sigma=0.1$. We set $\eta=10^{-4}$ for CVX solver and $\eta=10^{-2}$ for ADMM solver since CVX solver gets the exact solution while ADMM solver has performance degradation due to finite iteration. 
For each sampling probability, we sample the desired signals in time domain randomly and the results are averaged over 300 independent trials. Then we count the number of successful trials, and calculate the related probability. Here, we claim a trial as a successful trial if the solution $\vx^\dagger$ satisfies
$$
\frac{\norm{\vx^\star - \vx^\dagger}_2}{\norm{\vx^\star}_2} < \eta.
$$
The results are presented in Fig. \ref{fig: PerformanceComparison_002}. The results indicate that  the proposed approach \eqref{P: nuclear norm minimization with MC}  outperforms the standard Hankel matrix completion with reliable reference.

\begin{table}[t]
	\caption{Running time comparison for 1-D signals}
	\label{tab:runtime_1d}
	\centering
	\begin{tabular}{cccc}  
		\toprule
		Methods  & $16$ & $64$ & $96$ \\
		\midrule
		Proposed-CVX    & 0.483s  &   3.185s & 21.456s \\
		Proposed-ADMM   & 0.003s  &   0.017s & 0.028s \\
		\bottomrule
	\end{tabular}
\end{table}
We then compare the running time for the proposed method with different solvers when the dimension of signals is 16,64 and 96. The numerical simulations are carried on an Intel desktop with 2.5 GHz CPU and 8 GB RAM. The results in Table \ref{tab:runtime_1d} show that ADMM solver can dramatically improve the running time. Besides, the proposed scheme with ADMM solver has a much better performance compared with the standard Hankel matrix completion.

\subsection{Simulations for 2-D signals}

We proceed by giving the numerical results for two-dimensional signals.

Consider a  two-dimensional spectrally sparse signal $\vX^\star \in \R^{N_1 \times N_2}$ and the signal is a weighted superposition of $r$ complex sinusoids with unit amplitudes. The reference signal is created by $\vPhi=\vX^\star+\sigma \vN$, where the entries of the real and imaginary part of $\vN$ follow i.i.d. standard Gaussian distribution, i.e., $ \mbox{Re}(N_{ij}), \mbox{Im}(N_{ij}) \sim \mathcal{N}(0,1)$ for $i=1,\ldots,N_1, j=1,\ldots,N_2$.

We first show the recovery results for the proposed method and standard Hankel matrix completion. We set $N_1=10, N_2=10, r=3, m=20$ and $\sigma=0.1$. 2D-MUSIC is applied to obtain the location and amplitude of frequencies\cite{berger2010signal}. The results are presented in Fig. \ref{fig: PerformanceComparison_003}. The results show that the proposed method with different solvers can exactly recover the desired signals while Hankel matrix cannot.

We next present the successful reconstruction rate as a function of sampling probability standard Hankel matrix completion, the proposed method with CVX solver and the proposed method with ADMM solver. We set $N_1=10, N_2=10, r=3$ and $\sigma=0.1$. We increase the number of samples $m$ from 1 to 100. We Fig. 4 gives the simulation results. As expected, the proposed scheme with the CVX solver performs the best, followed by the proposed scheme with ADMM solver and standard Hankel matrix completion. 

\begin{table}[t]
	\caption{Running time comparison for 2-D signals}
	\label{tab:runtime_2d}
	\centering
	\begin{tabular}{cccc}  
		\toprule
		Methods  & $8\times 8$ & $10\times 10$ & $12\times 12$ \\
		\midrule
		Proposed-CVX       & 1.420s    & 3.449s  & 12.847s\\
		Proposed-ADMM     & 0.399s    & 0.529s & 0.786s \\
		\bottomrule
	\end{tabular}
\end{table}

Finally, Table \ref{tab:runtime_2d} compares the running time for the proposed method with different solvers when $N_1=N_2=8$,  $N_1=N_2=10$ and  $N_1=N_2=12$.  The results present that the proposed  scheme with ADMM solver has smaller running time than that with CVX solver, especially when the dimension of signals is large.

%

\section{Conclusion} \label{sec: Conclusion}
In this paper, we have integrated prior information to improve the performance of spectrally sparse signal recovery via structured matrix completion problem and have provided the related performance guarantees. Furthermore, we have designed corresponding ADMM algorithm to reduce the computational complexity. Both the theoretical and experimental results show that the proposed scheme outperforms standard Hankel matrix completion.   


\appendices
\section{Proof of Theorem \ref{thm:uniqueness}}

Dual certification is used to deviate the theoretical results. In particular, we use the golfing method from \cite{gross2011recovering} to proceed the process. And we adjust the methods from \cite{chen2014robust,ye2017compressive} to suit our model.

Recall the definition of the operator $\mathcal{A}_k: \mathbb{C}^{n_1 \times n_2} \to \mathbb{C}^{n_1 \times n_2}$ by
$$
\mathcal{A}_k(\vX) =  \ip{\vX}{\vA_k} \vA_k,\, k = 1,\ldots,n.
$$
Then each $\mathcal{A}_k$ is an orthogonal projection onto the one-dimensional subspace spanned by $\vA_k$.
The orthogonal projection onto the subspace spanned by $\{\vA_k\}_{k=1}^n$ is given as $\mathcal{A} = \sum_{k=1}^n \mathcal{A}_k$. Let $\mathcal{A}^\perp$ denote the orthogonal complement of $\mathcal{A}$.
The summation of the rank-1 projection operators in $\{\mathcal{A}_k\}_{k \in \Omega}$ is denoted by $\mathcal{A}_\Omega$,
i.e., $\mathcal{A}_\Omega \triangleq \sum_{k \in \Omega} \mathcal{A}_k$. Since $\Omega$ is a multi-set and there may exist repetitions in $\Omega$, $\mathcal{A}_\Omega$ may be not a projection operator.
The summation of distinct elements in $\{\mathcal{A}_k\}_{k \in \Omega}$ is denoted by $\mathcal{A}'_\Omega$, which is a valid orthogonal projection.

Before proving the theorem, let's review the proposed program
\begin{align*} 
&\min \limits_{\vz} ~~ \norm{\mathcal{H}(\vz)}_*-2\lambda \, \text{Re}\(\ip{\mathcal{G}(\vphi)}{\mathcal{H}(\vz)}\) \nonumber\\
&~\text{s.t.}~~~  \mathcal{P}_\Omega(\vz)= \mathcal{P}_\Omega({\vx}).
\end{align*}

We begin by presenting two lemmas, which are necessary for the proof.

\begin{lemma}[{\cite[Lemma 19]{ye2017compressive} \& \cite[Lemma 3]{chen2014robust}}] \label{lemma-Invertibility-PtWPt}Suppose that 
	\begin{equation*}
	\label{eq:rowcolsp}
	\sum_{j=1}^{n_2} \left(\sum_{i=1}^{n_1} |[\vA_k]_{i,j}|\right)^2 = 1,
	~ \text{and} ~
	\sum_{i=1}^{n_1} \left(\sum_{j=1}^{n_2} |[\vA_k]_{i,j}|\right)^2 = 1.
	\end{equation*}
	So we have 
	\begin{equation}
	\label{eq:incoherence2}
	\begin{aligned}
	\max_{1 \leq k \leq n} \norm{\vU^H \vA_k}_F^2  & \leq \frac{\mu r}{n_1}, \\
	\max_{1 \leq k \leq n} \norm{\vV^H \vA_k^H}_F^2 & \leq \frac{\mu r}{n_2}.
	\end{aligned}
	\end{equation}
	Then for any small constant $0<\epsilon\leq\frac{1}{2}$, one has
	\begin{equation}
	\norm{\mathcal{P}_{\mathcal{T}}\mathcal{A}\mathcal{P}_{\mathcal{T}}-\frac{n}{m}\mathcal{P}_{\mathcal{T}}\mathcal{A}_{\Omega}\mathcal{P}_{\mathcal{T}}}\leq\epsilon\label{eq:Invertibility-PtWPt}
	\end{equation}
	with probability exceeding $1-n^{-4}$, provided
	that $m>c\mu c_{\mathrm{s}}r\log n$ for
	some universal constant $c>0$ and $c_{\mathrm{s}}\triangleq\max\{\frac{n}{n_1},\frac{n}{n_2}\}$.
\end{lemma}

\begin{lemma}\label{lemma-Dual-Certificate}Consider a multi-set
	$\Omega$ that contains $m$ random indices. Suppose that the sampling
	operator $\mathcal{A}_{\Omega}$ obeys 
	\begin{equation} \label{eq:WellConditionPtAomegaPt}
	\left\Vert \mathcal{P}_{\mathcal{T}}\mathcal{A}\mathcal{P}_{\mathcal{T}}-\frac{n}{m}\mathcal{P}_{\mathcal{T}}\mathcal{A}_{\Omega}\mathcal{P}_{\mathcal{T}}\right\Vert \leq\frac{1}{2}.
	\end{equation}
	If there exists a matrix $\boldsymbol{W}$ satisfying 
	\begin{equation}
	\mathcal{A}'_{\Omega^{\perp}}\left(\boldsymbol{W}\right)=0,\label{eq:UV_W_Contained_in_AOmega}
	\end{equation}
	\begin{equation}
	\left\Vert \mathcal{P}_{\mathcal{T}}\left(\mbox{\upshape sgn}[\mathcal{H} (\vx^\star)]-\boldsymbol{W}-\lambda \mathcal{G}(\vphi) \right)\right\Vert _F\leq \frac{1}{7n}, \label{eq:W_component_T}
	\end{equation}
	and 
	\begin{equation}
	\left\Vert \mathcal{P}_{\mathcal{T}^{\perp}}\left(\boldsymbol{W} + \lambda \mathcal{G}(\vphi) \right)\right\Vert \leq\frac{1}{2},\label{eq:NormWTPerp}
	\end{equation}
	then the program (\ref{P: nuclear norm minimization with MC}) can achieve exact recovery, i.e., $\vx$ is the unique minimizer.
\end{lemma}
\begin{IEEEproof}
	See Appendix \ref{appdix:lemma-Dual-Certificate}.
\end{IEEEproof}

As shown in \cite{chen2014robust}, we generate $j_0$ independent random multi-sets $\{\Omega_{i}\}_{i=1}^{j_0}$ and each set contains $\frac{m}{j_0}$ entries. Note that the distribution of $\Omega$ and $\cup_{i=1}^{j_0}\Omega_{i}$ is the same.
Then we construct of a dual certificate $\boldsymbol{W}$ via the
golfing scheme:
\begin{enumerate}
	\item Define $\boldsymbol{F}_{0} \triangleq \mathcal{P}_{\mathcal{T}}\left(\mbox{\upshape sgn}[\mathcal{H} (\vx)]-\lambda \mathcal{G} (\vphi) \right)$;
	\item For every $i$ ($1\leq i\leq j_{0}$), set $$\boldsymbol{F}_{i} \triangleq \mathcal{P}_{\mathcal{T}}\left(\mathcal{A}-\frac{nj_0}{m}\mathcal{A}_{\Omega_{i}}\right)\mathcal{P}_{\mathcal{T}}\left(\boldsymbol{F}_{i-1}\right);$$
	\item Define $\boldsymbol{W} \triangleq \sum_{i=1}^{j_{0}}\left(\frac{nj_0}{m}\mathcal{A}_{\Omega_{i}}+\mathcal{A}^{\perp}\right)\left(\boldsymbol{F}_{i-1}\right)$.
\end{enumerate}
	
	By the construction, it's easy to see that $\vW$ is in the range space of $\mathcal{A}_\Omega \cup \mathcal{A}^\perp$, then
	\begin{equation} \label{eq:UV_W_Contained_in_AOmega}
		\mathcal{A}'_{\Omega^{\perp}}\left(\boldsymbol{W}\right)=0.
	\end{equation}
	
	By recursive calculation as \cite[Eq. (40)]{chen2014robust}, we can obtain 
	\begin{equation}
		-\mathcal{P}_{\mathcal{T}}\left(\boldsymbol{W}-\boldsymbol{F}_{0}\right)=\mathcal{P}_{\mathcal{T}}\left(\boldsymbol{F}_{j_{0}}\right).
	\end{equation}	
	Using Lemma \ref{lemma-Invertibility-PtWPt} yields	
	\begin{align}  \label{eq:PtW_UB}
	\left\Vert \mathcal{P}_{\mathcal{T}}\left(\boldsymbol{W}-\boldsymbol{F}_{0}\right)\right\Vert _{F} 
	&=\left\Vert \mathcal{P}_{\mathcal{T}}\left(\boldsymbol{F}_{j_{0}}\right)\right\Vert _{F}\nonumber \\
	&\le \norm{\mathcal{P}_{\mathcal{T}}\left(\mathcal{A}-\frac{nj_0}{m}\mathcal{A}_{\Omega_{i}}\right)\mathcal{P}_{\mathcal{T}}}^{j_{0}} \norm{\boldsymbol{F}_0}_F \nonumber\\
	& \leq \epsilon^{j_{0}} \norm{\boldsymbol{F}_0}_F 
	\leq \frac{1}{2^{j_0}} \norm{\boldsymbol{F}_0}_F.
	\end{align}
	
	Let 
	\begin{equation}
		j_0= \max\left\{\log \(7n \norm{\boldsymbol{F}_0}_F\),1\right\},
	\end{equation}
	then we have
	\begin{align}  \label{eq:PtW_UB}
	\left\Vert \mathcal{P}_{\mathcal{T}}\left(\boldsymbol{W}-\mbox{\upshape sgn}[\mathcal{H} (\vx^\star)]+\lambda \mathcal{G}(\vphi)\right)\right\Vert _{F} \leq 
	\frac{1}{7n}
	\end{align}
	except with a probability at most $j_0 n^{-4}=o(n^{-3})$, as long as
	$$m>c \mu c_{\mathrm{s}}r\log n  \max\left\{\log \(7n \norm{\boldsymbol{F}_0}_F\),1\right\}. $$

	For the last condition, using triangle's inequality yields
	\begin{equation*}
	\norm {\mathcal{P}_{\mathcal{T}^{\perp}}\left(\boldsymbol{W} + \lambda \mathcal{G}(\vphi) \right)} \leq \norm {\mathcal{P}_{\mathcal{T}^{\perp}}(\boldsymbol{W})}
	+ \norm{\mathcal{P}_{\mathcal{T}^{\perp}}(\lambda \mathcal{G}(\vphi))}.
	\end{equation*}
    According to the result of \cite[VI. E]{chen2014robust}, we have
    \begin{align*}
    	&\norm {\mathcal{P}_{\mathcal{T}^{\perp}}(\boldsymbol{W})}\\
    	&\le \sum_{l=1}^{j_{0}} \norm{ \mathcal{P}_{\mathcal{T}^{\perp}}\left(\frac{nj_0}{m}\mathcal{A}_{\Omega_{l}}+\mathcal{A}^{\perp}\right)\mathcal{P}_{\mathcal{T}}\left(\boldsymbol{F}_{l-1}\right)} \\ 
    	&\leq \sum_{l=1}^{j_{0}} \small\left(\frac{1}{2}\right)^{l-1}\left(\sqrt{\frac{nj_0\log n}{m}} \norm{ \boldsymbol{F}_{0}}_{\mathcal{A},2}+\frac{nj_0\log n}{m} \norm{ \boldsymbol{F}_{0}}_{\mathcal{A},\infty}\right)\\
    	&< \frac{2}{\Delta}(\norm{ \boldsymbol{F}_{0}}_{\mathcal{A},2}+\norm{ \boldsymbol{F}_{0}}_{\mathcal{A},\infty}), 
    \end{align*}
    as long as 
    \begin{multline*}
    m \ge c\max\{\mu c_{\mathrm{s}},\nu\} \\ \cdot \max\{\Delta^2,1\} \, r\log n  \max\left\{\log \(7n \norm{\boldsymbol{F}_0}_F\),1\right\},
    \end{multline*}
    where $\nu=o(\mu c_{\mathrm{s}} \log^2 n)$ from \cite[Appendix E]{ye2017compressive}. Set
	\begin{equation*}
		\Delta= \frac{4(\norm{ \boldsymbol{F}_{0}}_{\mathcal{A},2}+\norm{ \boldsymbol{F}_{0}}_{\mathcal{A},\infty})}{1-2 \norm{\mathcal{P}_{\mathcal{T}^{\perp}}(\lambda \mathcal{G}(\vphi))}}.
	\end{equation*}
	If $\norm{\mathcal{P}_{\mathcal{T}^{\perp}}(\lambda \mathcal{G}(\vphi))} < \frac{1}{2}$, we have	
	\begin{equation*}
	\left\Vert \mathcal{P}_{T^{\perp}}\left(\boldsymbol{W} + \lambda \mathcal{G}(\vphi) \right)\right\Vert \leq \frac{1}{2}.
	\end{equation*}
	Therefore, we conclude, if  
	\begin{multline*}
	m \ge \max\{\Delta^2,1\} \, c\mu c_{\mathrm{s}} r\log^3 n  \max\left\{\log \(7n \norm{\boldsymbol{F}_0}_F\),1\right\},
	\end{multline*}
	then with high probability, we can achieve the unique minimum.
	\begin{remark} Form the operator $\mathcal{H}_c(\vx)$ with wrap-around property, $\nu=o(\mu c_{\mathrm{s}})$ according to \cite[Appendix E]{ye2017compressive}. Therefore, we can get the following bound of sample size
		\begin{multline*}
		m \ge \max\{\Delta^2,1\} \, c\mu c_{\mathrm{s}} r\log n  \max\left\{\log \(7n \norm{\boldsymbol{F}_0}_F\),1\right\}.	
		\end{multline*}
	\end{remark}
\section{Proof of Lemma \ref{lemma-Dual-Certificate}} \label{appdix:lemma-Dual-Certificate}

Let $\vz = \vx + \vh$ be the minimizer to (\ref{P: nuclear norm minimization with MC}). We will show that $\mathcal{H} (\vh) = 0$. Then by the injectivity of the operator $\mathcal{H}$, we achieve $\vh = 0$, so we have $\vz = \vx$.

According to case 2 in the proof of  \cite[Lemma 20]{ye2017compressive},
$\norm{\mathcal{P}_{\mathcal{T}} (\mathcal{H} (\vh))}_F \geq 3 n \norm{\mathcal{P}_{\mathcal{T}^\perp} (\mathcal{H} (\vh))}_F$ leads to $\mathcal{H} (\vh) = 0$. So we only need to prove that when
\begin{equation}
\label{eq:case1}
\norm{\mathcal{P}_{\mathcal{T}} (\mathcal{H} (\vh))}_F \leq 3 n \norm{\mathcal{P}_{\mathcal{T}^\perp} (\mathcal{H} (\vh))}_F,
\end{equation}
we also have $\mathcal{H} (\vh) = 0$. In the subsequent analysis, we assume that the condition  \eqref{eq:case1} is correct.

According to the definition of nuclear norm, there exists $\vB$ such that $\ip{\vB}{\mathcal{P}_{\mathcal{T}^\perp} (\mathcal{H} (\vh))}=\norm{\mathcal{P}_{\mathcal{T}^\perp} (\mathcal{H} (\vh))}_*$ and $\norm{\vB} \le 1$. Then
$\mbox{\upshape sgn} (\mathcal{H} (\vx))+\mathcal{P}_{\mathcal{T}^\perp} (\vB)$ is a sub-gradient of the nuclear norm at $\mathcal{H} (\vx)$.
Then it follows that
\begin{equation}
\label{eq:pf_lemma_uniqueness:ineq1}
\begin{aligned}
&\norm{\mathcal{H} (\vx) + \mathcal{H} (\vh)}_* - 2\lambda \, \text{Re}\(\ip{\mathcal{G}(\vphi)}{\mathcal{H} (\vx)+\mathcal{H} (\vh)}\)
{} \\
& \geq \norm{\mathcal{H} (\vx)}_* -2\lambda \, \text{Re}\(\ip{\mathcal{G}(\vphi)}{\mathcal{H} (\vx)}\) \\
&\quad + 2\,\text{Re}\(\ip{\mbox{\upshape sgn}[\mathcal{H} (\vx)] 
+ \mathcal{P}_{\mathcal{T}^\perp} (\vB)- \lambda \mathcal{G}(\vphi)}{ \mathcal{H} (\vh)}\) \\
& = \norm{\mathcal{H} (\vx)}_* -2 \lambda \, \text{Re}\(\ip{\mathcal{G}(\vphi)}{\mathcal{H} (\vx)}\)
+ 2\, \text{Re}\(\ip {\vW} {\mathcal{H} (\vh)}\) \\
&\quad + 2\,\text{Re}\(\ip{\mbox{\upshape sgn}[\mathcal{H} (\vx)] + \mathcal{P}_{\mathcal{T}^\perp} (\vB)- \lambda \mathcal{G}(\vphi)-\vW}{ \mathcal{H} (\vh)}\).
\end{aligned}
\end{equation}
We can get $\text{Re}\(\ip {\vW} {\mathcal{H} (\vh)}\)= 0$ as shown in \cite[A.33-A.34]{ye2017compressive}. In addition, we have
\begin{align*}
\langle \mathcal{P}_{\mathcal{T}^\perp} (\vB), ~ \mathcal{H} (\vh) \rangle
{}  & = \langle \mathcal{P}_{\mathcal{T}^\perp} (\vB), ~ \mathcal{P}_{\mathcal{T}^\perp} (\mathcal{H} (\vh)) \rangle \\
{} & = \norm{\mathcal{P}_{\mathcal{T}^\perp} (\mathcal{H} (\vh))}_*.
\end{align*}
Then the inequality \eqref{eq:pf_lemma_uniqueness:ineq1} becomes
\begin{align}
\label{eq:pf_lemma_uniqueness:ineq2}
&\norm{\mathcal{H} (\vx) + \mathcal{H} (\vh)}_* - 2\lambda \, \text{Re}\(\ip{\mathcal{G}(\vphi)}{\mathcal{H} (\vx)+\mathcal{H} (\vh)}\) \nonumber\\
&\geq \norm{\mathcal{H} (\vx)}_* - 2 \lambda \, \text{Re}\(\ip{\mathcal{G}(\vphi)}{\mathcal{H} (\vx)}\)+ 2\norm{\mathcal{P}_{\mathcal{T}^\perp} (\mathcal{H} (\vh))}_* \nonumber\\
&\qquad \quad - 2\, \text{Re}\( \langle \vW - \mbox{\upshape sgn}[\mathcal{H} (\vx)] +\lambda \mathcal{G}(\vphi), \mathcal{H} (\vh) \rangle\).
\end{align}

Next, we are going to derive 
\begin{multline*}
	\norm{\mathcal{P}_{\mathcal{T}^\perp} (\mathcal{H} (\vh))}_* \\ - \text{Re}\(\ip {\vW - \mbox{\upshape sgn}[\mathcal{H} (\vx)] +\lambda \mathcal{G}(\vphi)} {\mathcal{H} (\vh)}\)\ge 0.
\end{multline*}
Noting that $\text{Re}\(x\)\le |x|$ for $x \in C$, it's enough to prove 
$$
\norm{\mathcal{P}_{\mathcal{T}^\perp} (\mathcal{H} (\vh))}_* - |\ip {\vW - \mbox{\upshape sgn}[\mathcal{H} (\vx)] +\lambda \mathcal{G}(\vphi)} {\mathcal{H} (\vh)}|\ge 0.
$$
By using the triangle inequality, we have 
\begin{align} \label{eq:pf_lemma_uniqueness:ineq3}
&\norm{\mathcal{P}_{\mathcal{T}^\perp} (\mathcal{H} (\vh))}_* - |\ip {\vW - \mbox{\upshape sgn}[\mathcal{H} (\vx)] +\lambda \mathcal{G}(\vphi)} {\mathcal{H} (\vh)}| \nonumber\\
 & \ge \norm{\mathcal{P}_{\mathcal{T}^\perp} (\mathcal{H} (\vh))}_* - |\langle \mathcal{P}_{\mathcal{T}^\perp} (\vW +\lambda \mathcal{G}(\vphi)), \mathcal{H} (\vh) \rangle | \nonumber\\
&\qquad \qquad
- |\langle \mathcal{P}_{\mathcal{T}} (\vW - \mbox{\upshape sgn}[\mathcal{H} (\vx)]+\lambda \mathcal{G}(\vphi)), \mathcal{H} (\vh) \rangle |.
\end{align}
Using Holder's inequality and the properties of $\vW$ (Eqs. \eqref{eq:W_component_T} and \eqref{eq:NormWTPerp}) yields
	\begin{align} \label{eq:pf_lemma_uniqueness:ineq4}
	&\norm{\mathcal{P}_{\mathcal{T}^\perp} (\mathcal{H} (\vh))}_* - |\ip {\vW - \mbox{\upshape sgn}[\mathcal{H} (\vx)] +\lambda \mathcal{G}(\vphi)} {\mathcal{H} (\vh)}| \nonumber\\
	{} &  \ge \norm{\mathcal{P}_{\mathcal{T}^\perp} (\mathcal{H} (\vh))}_* - \norm{\mathcal{P}_{\mathcal{T}^\perp} (\vW+\lambda \mathcal{G}(\vphi))} \norm{\mathcal{P}_{\mathcal{T}^\perp} (\mathcal{H} (\vh))}_*  \nonumber\\
	&\qquad \quad
	-\norm{\mathcal{P}_{\mathcal{T}} (\vW - \mbox{\upshape sgn}[\mathcal{H} (\vx)]+\lambda \mathcal{G}(\vphi))}_F \norm{\mathcal{P}_{\mathcal{T}} (\mathcal{H} (\vh))}_F \nonumber \\
	{} & \ge \norm{\mathcal{P}_{\mathcal{T}^\perp} (\mathcal{H} (\vh))}_*
	- \frac{1}{2} \norm{\mathcal{P}_{\mathcal{T}^\perp} (\mathcal{H} (\vh))}_* -\frac{1}{7 n} \norm{\mathcal{P}_{\mathcal{T}} (\mathcal{H} (\vh))}_F  \nonumber\\
	{}& \ge \frac{1}{2} \norm{\mathcal{P}_{\mathcal{T}^\perp} (\mathcal{H} (\vh))}_* -\frac{1}{7 n} \norm{\mathcal{P}_{\mathcal{T}} (\mathcal{H} (\vh))}_F.
	\end{align}
By using Eq. \eqref{eq:case1}, we obtain
\begin{align} \label{eq:pf_lemma_uniqueness:ineq5}
&\norm{\mathcal{P}_{\mathcal{T}^\perp} (\mathcal{H} (\vh))}_* - |\ip {\vW - \mbox{\upshape sgn}[\mathcal{H} (\vx)] +\lambda \mathcal{G}(\vphi)} {\mathcal{H} (\vh)}| \nonumber\\
{}& \ge \frac{1}{2} \norm{\mathcal{P}_{\mathcal{T}^\perp} (\mathcal{H} (\vh))}_* -\frac{3}{7} \norm{\mathcal{P}_{\mathcal{T}^\perp} (\mathcal{H} (\vh))}_* \nonumber\\
{}& =\frac{1}{14} \norm{\mathcal{P}_{\mathcal{T}^\perp} (\mathcal{H} (\vh))}_* \ge 0. 
\end{align}
Therefore, we get
\begin{align} \label{eq: pf_lemma_uniqueness_right}
&\norm{\mathcal{H} (\vx) + \mathcal{H} (\vh)}_* - 2\lambda \, \text{Re}\(\ip{\mathcal{G}(\vphi)}{\mathcal{H} (\vx)+\mathcal{H} (\vh)}\) \nonumber\\
&\geq \norm{\mathcal{H} (\vx)}_* - 2\lambda \, \text{Re}\(\ip{\mathcal{G}(\vphi)}{\mathcal{H} (\vx)}\)+\frac{1}{14} \norm{\mathcal{P}_{\mathcal{T}^\perp} (\mathcal{H} (\vh))}_*. 
\end{align}
Since $\vz = \vx + \vh$ be the minimizer to (\ref{P: nuclear norm minimization with MC}), we also have
\begin{align} \label{eq: pf_lemma_uniqueness_left}
\norm{\mathcal{H} (\vx) + \mathcal{H} (\vh)}_* - 2\lambda \, \text{Re}\(\ip{\mathcal{G}(\vphi)}{\mathcal{H} (\vx)+\mathcal{H} (\vh)}\) \nonumber\\
\leq \norm{\mathcal{H} (\vx)}_* - 2\lambda \, \text{Re}\(\ip{\mathcal{G}(\vphi)}{\mathcal{H} (\vx)}\).
\end{align}

Combining Eqs. \eqref{eq: pf_lemma_uniqueness_right} and \eqref{eq: pf_lemma_uniqueness_left}, we get $\mathcal{P}_{\mathcal{T}^\perp} (\mathcal{H} (\vh)) = 0$. By using \eqref{eq:case1}, we also have $\mathcal{P}_{\mathcal{T}} (\mathcal{H} (\vh)) = 0$. So we conclude $\mathcal{H} (\vh) = 0$ and $\vh=0$.


%

\bibliographystyle{IEEEtran}
\bibliography{IEEEabrv,HankelMatrixCompletion}

\end{document}